\documentclass[iop]{emulateapj}
\usepackage{epsfig}
\usepackage{color}
\usepackage{aas_macros}

\usepackage{amsfonts}
\usepackage{amssymb}
\usepackage{amsmath}
\usepackage{graphicx}
\usepackage{epsfig}
\usepackage{color}
\usepackage[colorlinks,linkcolor=blue,anchorcolor=green,citecolor=blue]{hyperref}
\usepackage{threeparttable}

\def\araa{ARA\&A}
\def\apj{ApJ}
\def\apjl{ApJ}
\def\apjs{ApJS}
\def\ao{Appl.~Opt.}

\def\aap{A\&A}

\def\mnras{MNRAS}

\def\prd{Phys.~Rev.~D}

\def\pasj{PASJ}

\def\nat{Natur}

\def\physrep{Phys.~Rep.}

\shorttitle{Delayed Energy Injection in GRB 170817A}
\shortauthors{Li et al. 2018}

\begin{document}

\title{Continued Brightening of the Afterglow of GW170817/GRB\,170817A as due to a Delayed Energy Injection}

\author{Bing Li\altaffilmark{1,2,3}, Long-Biao Li\altaffilmark{1,4}, Yong-Feng Huang\altaffilmark{1,4}, Jin-Jun Geng\altaffilmark{1,4},
Yong-Bo Yu\altaffilmark{1,4}, Li-Ming Song\altaffilmark{2,3}}

\altaffiltext{1}{School of Astronomy and Space Science, Nanjing University, Nanjing 210023, China; hyf@nju.edu.cn, gengjinjun@nju.edu.cn}
\altaffiltext{2}{Key Laboratory for Particle Astrophysics, Chinese Academy of Sciences, Beijing 100049, China}
\altaffiltext{3}{Particle Astrophysics Division, Institute of High Energy Physics, Chinese Academy of Sciences, China}
\altaffiltext{4}{Key Laboratory of Modern Astronomy and Astrophysics (Nanjing University), Ministry of Education, Nanjing 210023, China}

\begin{abstract}
The brightness of the multi-wavelength afterglow of GRB\,170817A is increasing unexpectedly even $\sim$ 160 days 
after the associated gravitational burst. Here we suggest that the brightening can be caused by a late-time energy 
injection process. We use an empirical expression to mimic the evolution of the injection luminosity, which is 
consisted of a power-law rising phase and a power-law decreasing phase. It is found that the power-law indices 
of the two phases are $0.92$ and $-2.8$, respectively, with the peak time of the injection being $\sim 110$ days. 
The energy injection could be due to some kind of accretion, with the total accreted mass being $\sim 0.006 M_\odot$. 
However, normal fall-back accretion, which usually lasts for a much shorter period, cannot provide a natural explanation. 
Our best-fit decaying index of $-2.8$ is also at odds with the expected value of $-5/3$ for normal fall-back 
accretion. Noting that the expansion velocities of the kilonova components associated with GW170817 is $0.1-0.3\,c$, 
we argue that there should also be some ejecta with correspondingly lower velocities during the coalescence 
of the double neutron star system. They are bound by the gravitational well of the remnant central compact 
object and might be accreted at a timescale of about 100 days, providing a reasonable explanation for the energy 
injection. Detailed studies on the long-lasting brightening of GRB\,170817A thus may provide useful information 
on the matter ejection during the merger process of binary neutron stars.
\end{abstract}


\section{Introduction}
\label{sect:intro}

The first binary neutron star (NS) merger event was witnessed when the gravitational wave (GW)
event GW170817 was detected by the advanced Laser Interferometer Gravitational-Wave Observatory (LIGO)
and the Virgo Interferometer on 17 August 2017 \citep{Abbott2017aPRL}.
About 1.7 s later, a faint short gamma-ray burst (GRB) GRB 170817A was detected by Fermi Gamma-Ray Telescope
and International Gamma-Ray Astrophysics Laboratory (INTEGRAL), which is considered to be associated
with GW170817 \citep{Abbott2017bApJ,Goldstein2017ApJ,Savchenko2017ApJ}.
About $\sim 11$ hours after the event, an optical counterpart was identified (a kilonova)  and named as AT2017gfo
\citep{Arcavi2017Nat,Coulter2017Sci,Drout2017Sci,Kasen2017Nat,Kasliwal2017Sci,Pian2017Nat,Smartt2017Nat,Soares2017ApJ,Valenti2017ApJ},
which verifies the hypothesis that $r$-process-induced kilonovae are associated with binary NS mergers
\citep{Lilx1998,Rosswog2005ApJ,Metzger2010MN,Metzger2017LRR}, and confirms that at least some short GRBs
originate from binary NS mergers (e.g., \citealt{Eichler1989Nat,Meszaros1992ApJ,Narayan1992ApJ,Oechslin2006MN,Nakar2011Nat},
and see \citealt{Berger2014ARAA} for a recent review).

The isotropic energy of GRB\,170817A is $E_{\rm iso} = (3.1 \pm 0.7)\times 10^{46} \rm~erg$,
and the isotropic peak luminosity is $L_{\rm iso} = (1.6 \pm 0.6) \times 10^{47} \rm~erg \, s^{-1}$.
These numbers are several orders of magnitude lower than those of typical short GRBs \citep{Abbott2017cApJ,Goldstein2017ApJ,Zou2017arX}.
It implies GRB 170817A may originate from an off-axis relativistic jet~\citep{Kathirga2018MN,Xiao17,Meng2018arX}
or a cocoon expanding at a mildly relativistic speed \citep{Abbott2017cApJ,Gottlieb2017arX,Lazzati2017ApJ,Murguia2017ApJ}.
In the standard GRB fireball model, a highly collimated relativistic outflow interacts with the
surrounding medium to produce broadband afterglow emission, which may be as luminous as
$L_{\rm X,iso} > 10^{44}$~erg~s$^{-1}$ in X-rays at early stages for an on-axis observer \citep{Racusin2011ApJ,Fong2015ApJ}.
However, for GRB\,170817A, the luminosity of the early X-ray afterglow is quite faint
and is far less than $ 10^{40}~\rm erg \, s^{-1}$ \citep{Evans2017Sci}. Continued monitoring reveals that
the X-ray afterglow began to rise to an observable lever $\sim 10$ days post-merger
\citep{Troja2017Nat}, and then the radio afterglow also became observable a week later \citep{Hallinan2017Sci}.
These observational features are highly consistent with a short GRB seen off-axis.

Recent observations with $Chandra$ revealed that the X-ray afterglow continued to increase unexpectedly
till $\sim$ 110 days post merger \citep{Margutti2017ApJ,Troja2017Nat,Ruan2018ApJ}.
\cite{Ruan2018ApJ} further found that the radio emission brightened at approximately the same rate as
that of the X-ray emission, which indicates that they share a common origin (also see: \citealt{Mooley2018Nat}).
Later, \citet{Lyman2018arX} reported that the late-time optical afterglow of GRB\,170817A
was continuously brightening till 110 days after the merger.
\citet{DAvanzo18arX} claimed that the X-ray and optical emission started to decrease $\sim 135$ days post-merger
according to their observations with $XMM--Newton$ and Hubble Space Telescope ($HST$).
However the X-ray brightness seems to be still rising even $\sim$ 153 -- 164 days later
as hinted by the latest data from $Chandra$ \citep{Troja2018ATel},
which was marginally consistent with the tendency of the radio data \citep{Resmi2018arX}.
Most recently, \citet{Dobie2018arX} reported that the radio afterglow seems to peak
at $\sim 150$ days post-merger and it now begins to decay.

The observed long-term brightening of the multi-band afterglow disfavors the simple off-axis top-hat jet
model \citep{Lazzati2017arX,Lyman2018arX,Margutti18arX}.
It has been proposed that the structured jet model or the cocoon model could match the late-time behavior of GRB\,170817A
\citep{Gottlieb2017arX,Granot2017arX,Lazzati2017arX,Lamb2017MN,Wang2017ApJ,Lyman2018arX,Mooley2018Nat,Troja2018arX}.
A structured outflow with a highly relativistic inner core launched by the merger remnant may interpret the
rising afterglow emission \citep{Granot2003ApJ,Kumar2003ApJ,Perna2003ApJ,Lazzati2004AA,Starling2005MN,Rezzolla2011ApJ}.
Alternatively, the brightening afterglow could arise from a wide-angle cocoon that breaks out at a mildly relativistic speed.
Cocoons are widely believed to be involved in the collapsar model for long GRBs
~\citep{Meszaros2001ApJ,Ramirez2002MN,Lazzati2005ApJ,Pe'er2006ApJ}.
During the NS-NS merger process, a jet-cocoon structure could also be produced
~\citep{Rosswog2013RSP,Hotokezaka2013PRD,Nagakura2014ApJ,Murguia2014ApJ}.
In this case, the late afterglow emission is generated from the interaction between the cocoon and
the interstellar medium~\citep{Nakar2010ApJ,Duffell2015ApJ,Rezzolla2015ApJ,Nakar2017ApJ,Gottlieb2018MN}.
Additionally, a jet-less isotropic fireball expanding ahead of the kilonova ejecta may also interpret the
observed brightening of GRB\,170817A before $\sim 100$ days \citep{Salafia2017AA,Salafia2018MN}.

Except for the models mentioned above, it is also possible that the brightening features are caused by a
continuous energy injection from the central engine.
Energy injections can be due to various activities of the remnant central compact object after the merger.
For example, the spinning down of a millisecond NS can supply enough energy for the external shock \citep{Geng2018arX}.
The accretion of circum-burst matter is another possibility \citep{Murase2018ApJ}.
In fact, accretion of the fall-back material is a quite common process in many circumstances \citep[e.g.,][]{Kumar2008MN,Cannizzo2011ApJ}.
In a few previous studies on other GRBs, fall-back accretion by BHs has been extensively
discussed \citep{Perna2006ApJ,Rosswog2007MN,Kumar2008MN,Zhang2008ApJ,Lee2009ApJ,Rossi2009MN}.
It can interpret some non-standard afterglow signatures, such as the X-ray plateaus and the optical
bumps/rebrightenings \citep[e.g.,][]{FanWei2005MN,WuXF2013ApJ,Geng2013ApJ,Laskar2015ApJ,YuHuang2015MN,Geng2016AdA}.
For GRB\,170817A, if the central engine was restarted and capable of releasing energy over a long time,
then the long-last brightening of the afterglow could be explained.

Inspired by these works, we investigate the possibility that the continued brightening of the
afterglow of GRB\,170817A is due to some kind of energy injection from the central engine.
We assume that the energy injection is consisted of a rising phase and a decreasing phase.
The injection luminosity in each phase is taken as a power-law function of time.
We compare our modeling results with the observations to derive the required characteristic injection
luminosity and the power-law indices, and then discuss possible theoretical implications of the results.
Our article is organized as follows. In Section 2, the energy injection process is briefly introduced.
Numerical results on the multi-band afterglow light curves are presented and compared with the
observational data in Section 3. Finally, we discuss and summarize our results in Section 4.

\section{Model}

Energy injections can be induced by various mechanisms, such as the spinning down of
rapidly rotating neutron stars or some kinds of accretion. In different situations, the injection
luminosity should also differ markedly. Among these processes, of special
interests is the fall-back accretion of materials around newly born compact stars.
For example, in the collapsar model for long GRBs, numerical simulations show that a significant portion of
the ejected material of the extended stellar envelope will fail to escape.
They will continuously fall-back toward the central remnant, resulting in a long-lasting
reactivation of the central engine \citep{Kumar2008Sci}. Similarly,
fall-back process has also been revealed by various simulations of double NS mergers \citep{Rosswog2007MN,Hotokezaka2013PRD}.
Hinted by these effects, we will use an empirical expression to approximate the
varying luminosity of the energy injection in our study.
We then investigate the dynamics of the external shock under such an pointing-flux dominated injection,
which should naturally deviate from that in the standard fireball model \citep{Panaitescu1998ApJ}.

\subsection{Two phases of the Energy Injection}

After the merger of a binary NS system, the central remnant could be either a magnetized NS or a
promptly-formed black hole (BH), depending on the masses of the two NSs and the equation of state
of supranuclear matter \citep{Hotokezaka2013ApJ,Piro17}.
The remnant of GW170817 is still under debate \citep{Shibata2017PRD}.
\citet{Pooley2017arX} argued that the early X-ray data of GW170817 are better explained by a BH,
rather than a hyper-massive neutron star. By considering the upper limits on the electromagnetic energy release
and the ejecta mass during the binary NS merger of GW170817, \citet{Margalit2017ApJ} examined various
equations of state for neutron stars, and derived the maximum
mass limit of NS as 2.17 $M_\odot$.
A BH as the remnant of the merger is thus strongly indicated since
the total mass of the two neutron stars as estimated from GW observations is significantly larger than
the mass limit. Finally, the $1.7$~s delay between the GRB and the GW signal may also support a BH born at the center.
Here, for simplicity, we assume that a BH is formed at the center after GW170817.
So, we do not consider the dipolar radiation from a rapidly rotating neutron star as the
source for energy injection. Instead, we consider the energy extraction from the BH via some accretion process.

For a rotating BH with an accretion disk, the rotation energy could be extracted through the Blandford-Znajek mechanism
\citep{Blandford1977MN,Ghosh1997MN,Livio1999ApJ,Lee2000ApJ,Lee2000PhR,Zhang2002ApJ},
which results in a Poynting-flux dominated outflow.
The detailed luminosity profile for the Poynting-flux depends on the exact accretion process.
For a fall-back like accretion, the accretion rate may initially increase with time \citep{MacFadyen2001ApJ}.
After reaching a peak value, the accretion rate decreases with time as
$\dot{M} \varpropto t^{-5/3}$ \citep{Rees1988Nat,Chevalier1989ApJ,MacFadyen2001ApJ,Rosswog2007MN}.
In the case of GW170817, the accretion surely is not a constant process. The injection power should depend
on many parameters such as the mass ($M_{\rm BH}$) and spin ($a$) of the BH, the density and velocity distribution
of the circum-burst medium, and even the surrounding magnetic field \citep{Zhang2008ApJ,Dai2012ApJ,WuXF2013ApJ,YuHuang2015MN,ChenW2017ApJ}.
As a result, it is reasonable to assume that the injection luminosity profile should
be consisted of two phases, a rising phase and a decreasing phase.
Following the work of \cite{Geng2013ApJ}, we take the injection luminosity profile
as a broken-power-law function (jointing the rising phase and the decreasing phase smoothly), i.e.,
\begin{equation}
L = L_{\rm p} [ \frac{1}{2} (\frac{t}{t_{\rm p} })^{- \alpha_r s} + \frac{1}{2} (\frac{ t}{t_{\rm p} })^{- \alpha_d s}]^{-1/s},
\end{equation}
where $L_{\rm p}$ is the peak luminosity at the peak time $t_{\rm p}$,
$\alpha_{r}$ and $\alpha_{d}$ are respectively the rising and decreasing index,
and $s$ represents the sharpness of the peak.
Figure 1 shows the injection luminosity profile used in our modeling.

\subsection{Shock Dynamics}

In our work, we use the generic dynamical equations to calculate the evolution of the GRB outflow
\citep{Huang1999MN,Huang2000MNR}, which can be conveniently used to calculate the afterglow light
curves under various physical conditions \citep[e.g.,][]{HuangYF2006ApJ,KongSW2010MN,Geng2014ApJ,LilongB2015MN}.
However, note that when there is an energy injection of Poynting-flux, the dynamical equation for GRB ejecta should be modified.

Let us consider a GRB ejecta with the initial mass of $M_{\rm ej}$, the bulk Lorentz factor
of $\Gamma$ ($\Gamma=1/\sqrt{1-\beta^{2}}$, with $\beta c$ being its velocity), and a half-opening angle of $\theta_j$.
The evolution of the bulk Lorentz factor of the forward shock can then be described by \citep[e.g.][]{Liu2010SCP,Geng2013ApJ,Geng2016AdA},
\begin{equation}
\frac{d \Gamma}{d M} = - \frac{(\Gamma^{2} - 1)-\frac{1 - \beta}{\beta c^{3}} \Omega_{j} L(t_{\rm b} - R/c) \frac{d R}{d M}}{M_{\rm ej} + 2 (1 - \varepsilon) \Gamma M + \varepsilon M},
\end{equation}
where $t_b$ is the time measured in the burst frame,
$\Omega_{j} = (1 - \cos \theta_{j})/2$ is the beaming factor of the jet,
$\varepsilon$ is the radiative efficiency, and $m$ and $R$ are the swept-up mass
and radius of the shock, respectively.
According to Eq. (2), the delayed energy injection with a luminosity of $L$ can change the evolution of $\Gamma$,
thus the afterglow light curve should show a plateau or a re-brightening correspondingly.

\section{Numerical Results}

Considering the energy injection luminosity and the dynamical evolution as described in Section 2,
we now calculate the multi-band afterglow light curves for GW170817/GRB\,170817A.
In our calculations, the initial value of the bulk Lorentz factor is taken as $\Gamma_0=100$, a typical value
for GRB outflow \citep[e.g.,][]{Racusin2011ApJ}.
We fix the power-law index of the electron spectrum as $p=2.2$, which has been derived from the spectrum
of the brightening afterglow of GRB\,170817A by some researchers (see \citealt{Dobie2018arX,DAvanzo18arX,Margutti18arX,Ruan2018ApJ} for details).
According to the studies on the environment of short GRBs, the circum-burst density is generally low.
It is typically in the range of $\sim 10^{-3}$ -- $10^{-2}\,\rm cm^{-3}$ \citep{Fong2015ApJ, Hallinan2017Sci}.
We thus set the number density as $n = 1 \times 10^{-3}\,\rm cm^{-3}$ in our calculations.
Other parameters, such as $\theta_{\rm obs}, \theta_{j}, E_{\rm K,iso}, \epsilon_e, \epsilon_B, L_{\rm p},
t_{\rm p}, \alpha_r, \alpha_d$ and $s$ are free parameters. We derive them by comparing our theoretical light curves
with observational data. Our fit results are plotted in Figures 1~---~3, and the derived parameter
values are listed in Table 1. Generally, we find that our late-time energy injection model can well explain
the multi-band afterglow of GRB\,170817A.
The corresponding chi-square of the fit is $\chi^{2}\approx 42.3$, with 22 degrees of freedom.

Figure 1 shows the evolution of the luminosity of the energy injection in our modeling.
At early stages, the injection luminosity is a power-law function of time, $L(t) \propto t^{0.92}$.
Then at late times, the luminosity decays with time as $L(t) \propto t^{-2.8}$.
In Figure 2, we plot the bulk Lorentz factor versus time, where the dotted line corresponds to
a jet without energy injection and the solid line corresponds to the jet with energy injection.
From this figure, we see that the energy injection begins to affect the dynamics at about
$\sim 10$ days. It increases the Lorentz factor significantly, so that the relativistic phase lasts for
about $\sim 100$ days. Such a change in the dynamics will finally lead to a brightening in the multi-band afterglow,
as shown in Figure 3.
In our calculations, we basically ignored the lateral expansion of the ejecta.
To examine the effect of lateral expansion, we have also calculated the evolution of the
jet (also with the energy injection) by assuming that it expands laterally at the co-moving
sound speed. The result is plotted as the dashed line in Figure 2. We see that the
difference between the dashed line and the solid line is very small, showing that this effect
is not significant.

Figure 3 illustrates our fit result for the multi-band afterglow of GRB\,170817A.
We see that our model can well explain the multi-wavelength
observational data. The derived parameters are also in reasonable ranges (see Table 1).
In our calculations, an isotropic energy of $E_{K,\mathrm{iso}} \simeq 4.0\times 10^{51}$ erg is used, which is
comparable to that derived by \citet{Hallinan2017Sci}.
The values of other normal parameters of the external shock, such as $\theta_j$, $\theta_{\rm obs}$,
are all in reasonable ranges as compared with those adopted in previous afterglow
studies \citep{Lyman2018arX,Mandel2018ApJ,Margutti18arX}.
In our theoretical light curves, the peak time of the afterglow is around $200$ days.
However, note that the exact peak time is actually mainly determined by the
peak time of the energy injection ($t_p$). This parameter thus could hopefully give us helpful
information on the nature of the energy injection.

\begin{table}
\centering
\caption{Parameters used in fitting the afterglow of GRB\,170817A.\label{table1}}
\begin{tabular}{ccc}
\hline
\hline
Fireball parameter                          & value                \\
\hline
$\theta_{\rm obs}$ (degree)\tnote{a}        & 23.4                   \\
$\theta_j$ (degree)                         & 11.0                   \\
$E_{K,\mathrm{iso}}$ (erg) \tnote{b}        & $4.0 \times 10^{51}$   \\
$\epsilon_{e}$                              & 0.025                  \\
$\epsilon_{B}$                              & 0.0045                 \\
\hline
Injection Parameter                         & value  \\
\hline
$L_{\rm p}$ (erg s$^{-1}$)                  & $3.2 \times 10^{44}$   \\
$t_{\rm p}$ (days)                          & 110.0                  \\
$\alpha_{r}$                                & 0.92                   \\
$\alpha_{d}$                                & -2.8                   \\
$s$                                         & 0.38                   \\
\hline
\end{tabular}
\begin{tablenotes}
\item[a] $\theta_{\rm obs}$ is defined as the angle between the line of sight and the jet axis.
\item[b] $E_{\rm K,iso}$ is the initial isotropic kinetic energy of the ejecta.
\end{tablenotes}
\end{table}

\begin{figure}
 \centering
 \includegraphics[width=0.5\textwidth]{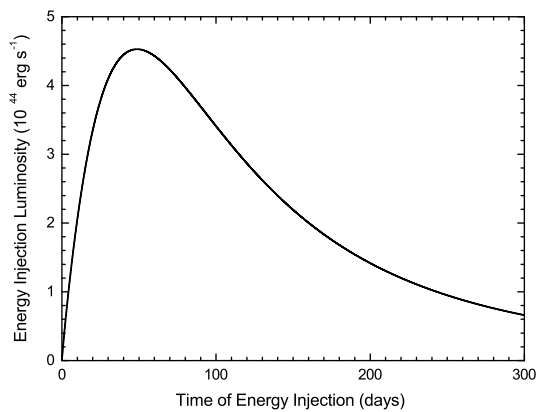}
 \caption{The evolution of the injection power. The luminosity profile is
          adopted as a broken power-law function (see Equation (1)) in our modeling.
          Detailed parameters are listed in Table 1.}
    \label{fig1}
\end{figure}

\begin{figure}
 \centering
 \includegraphics[width=0.5\textwidth]{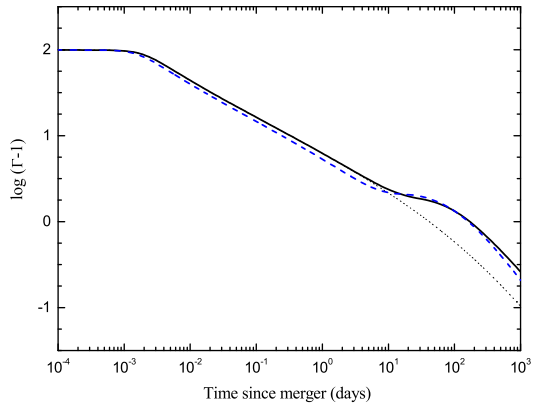}
 \caption{The evolution of the bulk Lorentz factor ($\Gamma$) of the ejecta.
          The thick solid line corresponds to a jet with energy injection and
          the thin dotted line shows the case of a normal jet without any energy injections.
          Note that in both cases, the jets do not expand laterally in our calculations.
          To show the effect of lateral expansion, the thick dashed line corresponds to a
          jet (with energy injection) that expands laterally at the co-moving sound speed.
          Other parameters involved in the calculations are listed in Table 1.}
    \label{fig2}
\end{figure}

\begin{figure}
 \centering
 \includegraphics[width=0.5\textwidth]{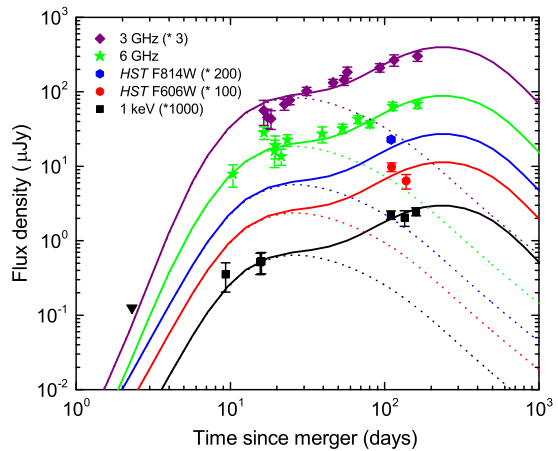}
 \caption{Our theoretical afterglow light curves as compared with the multi-band observations of GRB\,170817A.
        The solid lines represent the emission from the off-axis jet with the energy injection.
        As a comparison, the dotted lines show the same off-axis jet emission without any energy injections.
        The radio data points are taken from~\cite{Alexander2017},~\cite{Hallinan2017Sci},~\cite{Kim2017ApJ},
        \cite{Mooley2018Nat}, and \cite{Margutti18arX}.
        The X-ray data points are taken from~\cite{Ruan2018ApJ}, ~\cite{Lazzati2017arX}, and~\cite{DAvanzo18arX}.
        The optical (in {\it HST} optical bands of F814W and F606W) data points are taken from
        \cite{Lyman2018arX} and~\cite{Margutti18arX}.
        Note that the 1 keV upper limit shown as the black triangle is taken from \citet{Margutti2017ApJ}.}
    \label{fig3}
\end{figure}

From the injection luminosity profile as illustrated in Figure 1, we can estimate the
total accreted mass in the process.
In principle, assuming that the potential energy of the accreted material is converted to the jet power
at a particular ratio of $\eta$, we have
\begin{equation}
\frac{\eta G M_{\rm BH} M_{\rm fb}}{r_{\rm disc}}  =
\frac{(1 - \cos\theta_j)}{2} \frac{1} {1+z} \int_{}^{}L(t) \rm dt,
\end{equation}
where  $r_{\rm disc}$ is the radius of the accretion disc around the BH
and is taken as ten times that of the Schwarchild radius ($2 G M_{\rm BH}/c^2$).
Therefore, a total accreted mass of $M_{\rm fb} \simeq 0.006 M_{\odot} (\eta / 0.1)^{-1} (M_{\rm BH} / 2.7 M_{\odot})^{-2/3}$
is needed in our scenario. This mass is not too large. It could come from the ejected material during
the double NS merger process. As a comparison, the mass of the dynamical ejecta of
double NS mergers is typically $\sim 10^{-4}$--$10^{-2} M_{\odot}$ \citep{Rosswog2007MN,Hotokezaka2013PRD,Bauswein13}.
It also does not conflict with the estimation of $M_{ej} \sim 10^{-3}$~--~$10^{-2} M_{\odot}$ for the dynamical
ejecta of GW170817 \citep{Abbott2017dApJ,Utsumi2017PASJ,Tanaka2017PASJ,Matsumoto2018arX}.
In a few recent studies, a more precise ejecta mass of $M_{ej} \sim 0.05 M_{\odot}$ was estimated when modelling
the GW170817 kilonova \citep{Cowper2017ApJ,Drout2017Sci,Kasliwal2017Sci,Kasen2017Nat,Villar2017ApJ,Waxman2017arX}.
These studies also support the idea that there is enough material being ejected during the NS-NS merger.

\section{Discussion and Conclusion}

The X-ray, optical and radio emission of the afterglow of GRB\,170817A shows a steady rise
till $\sim 160$ days after the merger.
In this study, we suggest that the afterglow is
produced by an off-axis jet with a delayed energy injection caused by some kind of accretion.
By fitting the observed multi-band light curves, we derived the required injection luminosity
profile, which is composed of a power-law rising phase and a power-law decreasing phase.
We show that our model can well explain the X-ray, optical and radio light curves of GRB\,170817A.

In our modeling, the peak time of the energy injection is about 110 days, and a total
accreted mass of $\sim 6 \times 10^{-3} M_{\odot}$ is needed for the energy injection.
It is thus a key issue that how such an accretion can be achieved.
A possible solution is the so called fall-back accretion.
The fall-back accretion process was firstly discussed by \citet{Colgate1971ApJ}.
It is originally assumed to be related to supernovae. Later, people realized that fall-back
accretion could also play an important role in the collapsar model for long GRBs
\citep{Chevalier1989ApJ,Kumar2008Sci,Zhang2008ApJ,WuXF2013ApJ,Geng2013ApJ}.
It is interesting to note that evidence for fall-back accretion during binary NS mergers
that produce short GRBs has also been hinted.
In fact, for short GRBs, even if a small mass of $\sim 10^{-5}$ -- $10^{-4}M_{\odot}$
is accreted after the NS-NS merger, it would be sufficient to power a bright afterglow \citep{Metzger2010MN}.
Some possible examples include short GRB\,060729~\citep{Cannizzo2011ApJ}
and short GRB\,130603B~\citep{Hotokezaka2013ApJ,Kisaka2015ApJ}. It is possible that fall-back accretion and
related energy injection may be a common process among short GRBs.

However, for the brightening of the afterglow of GRB\,170817A, normal fall-back accretion maybe cannot
provide a natural explanation, since our best-fit parameters seem at odds with a fall-back interpretation.
In general, during the double NS merger process, one would expect the fall-back accretion to peak on
about the dynamical timescales of the least bound material, which would be less than $\sim 100$ millisecond or so.
More detailed investigation also shows that the normal fall-back accretion usually does not last for a
long period~\citep{Rosswog2007MN}. Additionally, our best fit slope for the late-time segment of the
luminosity profile is $\alpha_{\rm d} \sim -2.8$. This value is also inconsistent with the value of $-5/3$ expected for a
normal fall-back accretion. We thus need to seek for a mechanism other than normal fall-back accretion
for the energy injection in GRB\,170817A.

We note that a bright kilonova has been found to be associated with GW170817.
The kilonova is composed of ``blue'', ``purple'' and ``red'' components.
Their expanding velocities are typically in the range of $0.1c$ -- $0.3c$,
and their total mass can be as large as $\sim 0.08 M_\odot$ \citep{Villar2017ApJ}.
The material of these kilonova components undoubtedly should be ejected from the binary NS merger.
But since their velocities are too large, they could not be accreted during the afterglow stage.
However, we could imagine that there should also be many slower materials ejected during the process.
Their velocities are significantly higher than those of the initial fall-back materials, but they
are still bound by the gravity of the central BH. They could be
accreted at a much later stage, i.e. about 110 days later, providing a source for the late energy
injection as required in our modeling. Note that according to our calculations,
a total accreted mass of $\sim 6 \times 10^{-3} M_{\odot}$ is needed for the energy injection.
This value is much less than the total mass of the observed kilonova components.
It is also consistent with the typical mass range of the dynamical ejecta estimated by a few other
authors ($\sim 10^{-4} - 10^{-2} M_{\odot}$,
\citealt{Hotokezaka2013PRD,Rosswog2013RSP,Just2015MN,Rosswog2007MN,Rossi2009MN,Kisaka2015ApJ}).

Energy extraction from a rotating BH through the
Blandford-Znajek mechanism is usually assumed to be in the form of a Poynting flux.
In this case, macroscopical magnetic field may present in the external shock
and may play an important role in our energy injection model. A possible consequence
is that the induced afterglow emission is likely to be polarized
to some extent.
However, it is also possible that the injected energy does not necessarily be in the form of a
Poynting flux in realistic case. It may simply re-energize the external shock and increase the
afterglow brightness. Consequently, no polarization will be expected in the brightening stage.
Polarization observations can help to discriminate between these two cases.
More observations on the late time afterglows of short GRBs should be conducted in the future.

In our calculations, although we have assumed that the central engine of GRB\,170817A is a BH,
an NS as the central remnant still cannot be ruled out. A newly born NS accompanied
by an accretion disk may also play a similar role to inject energy into the
external shock \citep{Dai2012ApJ,ZhangD08}. To produce a comparable
brightness, the total accreted mass may need to be slightly larger to ensure a similar
potential energy release.

\acknowledgments
We thank the anonymous referee for valuable comments and suggestions that lead to an
overall improvement of this study.
We also would like to thank Wei-Hua Lei, Ling-Jun Wang and Qiang Zhang for helpful discussion.
This work is supported by the National Natural Science Foundation of China (Grants No. 11473012),
the National Basic Research Program of China (``973'' Program, Grant No. 2014CB845800),
the National Postdoctoral Program for Innovative Talents (Grant No. BX201700115),
China Postdoctoral Science Foundation funded project (Grant No. 2017M620199),
and by the Strategic Priority Research Program of the Chinese Academy of Sciences
``Multi-waveband Gravitational Wave Universe'' (Grant No. XDB23040000 and XDB23040400).
LMS acknowledges support from the National Program on Key Research and Development
Project (Grant No. 2016YFA0400801) and the National Basic Research Program of China
(Grant No. 2014CB845802).

\clearpage

\end{document}